\documentclass[aps,prb,reprint,amsmath,amssymb,superscriptaddress,showpacs,floatfix]{revtex4-1}

\usepackage{graphicx}
\usepackage{dcolumn}
\usepackage{bm}
\usepackage[usenames, dvipsnames]{color}
\usepackage{multirow}

\begin{document}

\title{Anderson transitions in disordered two-dimensional lattices}
\author{Dayasindhu Dey}
\email{dayasindhu.dey@bose.res.in}
\affiliation{S. N. Bose National Centre for Basic Sciences, Block - JD, Sector - III, Salt Lake, Kolkata - 700098, India}
\author{Manoranjan Kumar}
\email{manoranjan.kumar@bose.res.in}
\affiliation{S. N. Bose National Centre for Basic Sciences, Block - JD, Sector - III, Salt Lake, Kolkata - 700098, India}
\author{Pragya Shukla}
\affiliation{Department of Physics, Indian Institute of Technology Kharagpur,
             Kharagpur - 721302, India.}

\date{\today}

\begin{abstract}

We numerically analyze the energy level statistics of the
Anderson model with Gaussian site disorder and constant hopping.
The model is realized on  different two-dimensional
lattices, namely, the honeycomb, the kagom\'e, the square, and 
the triangular lattice. By calculating
the well-known statistical measures viz., nearest neighbor spacing 
distribution, number variance, the partition number  and the dc 
electrical conductivity from Kubo-Greenwood formula, we show that there is 
clearly a delocalization to localization transition with increasing 
disorder. Though the statistics in different lattice systems differs 
when compared with respect to the change in the disorder strength only, 
we find there exists a single complexity parameter, a function of 
the disorder strength, coordination number, localization length, and the
local mean level spacing, in terms of which the statistics of the
fluctuations matches for all  lattice systems at least
when the Fermi energy is selected from the bulk of the energy levels.
\end{abstract}


\maketitle

\section{\label{intro}Introduction}
The effect of impurities in the low dimensional systems such as Graphene 
\cite{geim2004, geim2007, geim2009, geng2012, beenakker2007}, 
nano flakes \cite{romer2013}, metallic thin films \cite{biswas2014, liao2015},arrays of 
quantum dots etc., has been an intense  area of research in the last 
decade. Some of these low dimensional structures have potential 
applications in electronic devices, because of their finite size and the  
performance under varying degree of disorder and dimensionality \cite{liao2015, lu2011}.  
In disordered systems delocalization to localization transition is 
one of the intersting phenomenon, therefore to caracterize this transition,
analysis of the electronic wavefunction is necessary. 
The extended to localized transition, also known as  the `Anderson transition',  
and its dependence on  various system parameters such as,  disorder,  
dimensionality, lattice structure, system size etc., in the finite systems are  
a frontier area of research.  This motivates us to seek an estimation of 
the critical disorder for the transition in finite size lattices 
with different geometry.

In 1958, Anderson suggested that an electron inside a material 
can be fully localized in the presence of a large disorder 
\cite{anderson58},  whereas Edwards {\it et al.} showed that a  transition from 
an extended to a localized states in the square lattice can occur at a 
disorder strength of 5 or 6 times higher than the band width of the 
Anderson model Hamiltonian \cite{thouless72}.  
In 1979, Abrahams {\it et al.} conjectured, based on a scaling hypothesis, 
that the electronic states are localized  in less than three 
dimensional (3D) systems in the presence of any amount of disorder 
\cite{abrahams79,lee85}. The scaling hypothesis was later on  supported 
by many studies of the  Anderson model  on square lattices 
\cite{mackinnon81, *mackinnon83, *mackinnon85, *pichard81a, *pichard81b}. 
Further, Altshuler  {\it et al.} showed that weak electron interaction enhances 
the localization in these systems, whereas under strong electron 
interaction, 2D electrons behave like a Wigner crystal and, as 
 shown by  Tanatar {\it et al.} \cite{altshuler80}, even a small 
amount of disorder can make the system insulating at zero temperature. 
The theoretical analysis \cite{abrahams2001} suggested that under strong electron 
interactions and small disordered regime, 2D systems can   be conducting. 
Most of these studies are done for infinite systems and square lattices, except some of the 
recent studies on the honeycomb lattices 
\cite{ostrovsky2007, *ryu2007, *nomura2007, *san-jose2007, *lewenkopf2008, *beenakker2007}.
A detailed review on the Anderson transition is given in the Ref.~\onlinecite{evers2008}. 

The study of Anderson transition  is not only important for material 
sciences but also relevant  to understand the influence of the 
wavefunction dynamics on the physical properties of the disordered systems. 
The presence of disorder and/or interaction leads to a randomization of 
the Hamiltonian, resulting in a random  matrix  representation in a 
physically suitable basis e.g., site basis. The structure of the matrix 
e.g., degree of sparsity is sensitive to various system conditions viz., 
dimensionality, shape, size, and boundary conditions. The 
statistical behaviour of the system can therefore be analysed by an 
ensemble of the disordered Hamiltonians.  
Such analysis has been a subject of extensive study during the past decade. 
It is now well-known that, in the weak disorder regime, the statistics 
can be well-modelled by the Wigner-Dyson universality classes of random 
matrix ensembles which correspond to extended, featureless eigenfunctions 
and a strong level-repulsion with  statistics of the eigenvalues  and the  
eigenfunction independent from each other. Increasing the disorder in 
finite size systems causes  the statistics to crossover from  Wigner-Dyson 
to the Poisson universality class (with no level-repulsion and fully 
localized eigenfunctions in strong disorder limit). The statistics in  the 
intermediate regime e.g., near critical disorder is sensitive to the degree 
of eigenfunction localization which in turn is expected to depend on the 
system conditions besides disorder. The study \cite{mirlin2k,evers2008} 
however, showed that the statistics can be well-modeled by the single 
parametric power-law random banded matrix (PRBM) ensembles. 
Another theoretical study \cite{shukla05} later on indicated the application 
of a wide range of random matrix ensembles (besides PRBM) as the model for 
the intermediate statistics; this study  was  based on the common 
mathematical formulation of the energy level statistics of a broad class of 
random matrix ensembles (with varying degree of sparsity and disorder but 
same symmetry class). The formulation  is governed by a single parameter, 
referred as the complexity parameter, a function of all system parameters 
including energy range of interest and therefore different ensembles are 
expected to show analogous statistics if their complexity parameters are 
same \cite{shukla05}.  As the theoretical claim about the existence of a 
complexity parameter is in clear agreement with the single parameter 
scaling conjecture of Ref.~\onlinecite{abrahams79}, it is highly desirable to 
seek its numerical validity in disordered systems.  
\begin{figure}[!t]
\includegraphics[width=\columnwidth]{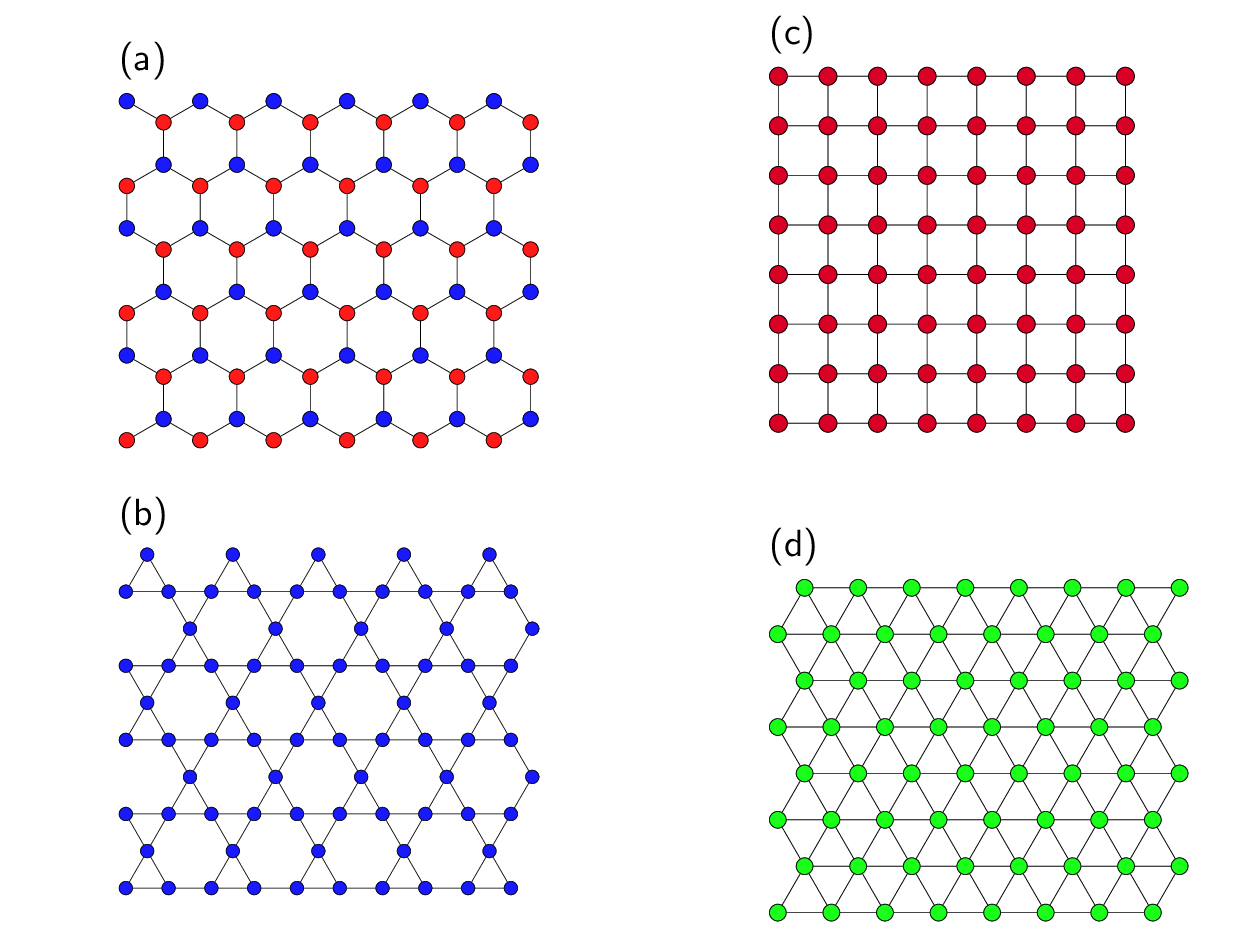}
\caption{\label{fig:lattices}(Color online) Two dimensional
lattice structures  considered in this paper: (a) the honeycomb lattice 
with coordination number 3, (b) the kagom\'e and (c) the square lattices 
both having 4 nearest neighbors  and (d) the triangular lattice with 
coordination number 6}
\end{figure}

In  this paper, we  study the  transitions from extended to localized 
state of Anderson Hamiltonian on finite 2D systems with different 
geometry in presence of the disorder. As shown in Fig.~\ref{fig:lattices}, 
we consider four lattice geometries, namely,  square, triangular, 
honeycomb and kagom\'e lattice which differ from each other in 
the coordination number and their bond connectivity. The coordination numbers 
of honeycomb, square, kagom\'e  and triangular,  lattice are   three, 
four, four, and six respectively.  We note that the square and the 
kagom\'e both have four nearest neighbour but their spectrum is 
completely different because of the bond connectivity.  
The paper is organized as follows. In section~\ref{sect2}, 
we introduce the model disordered Hamiltonian; here the 
theory leading to single parameter based formulation of 
the spectral statistics is also briefly reviewed. 
The section~\ref{sec:fluc} briefly describes 
the statistical measures used in our numerical analysis. 
The section~\ref{sec:res} presents the details of the numerical techniques 
as well as  results; this  section is 
divided into four sub-sections one for each 
lattice system. The comparison of the statistics of physical parameters
with respect to the single complexity parameter for various lattices 
is given in  section~\ref{sect5}. In section~\ref{sect6}, results are 
discussed and compared with the existing literature.

\section{\label{sect2} The Model }
{\it The Hamiltonian}: We consider the standard Anderson model \cite{anderson58}  
under tight binding approximation realized on four different 
2D lattice systems.  The tight binding  model Hamiltonian 
is well suited for a metallic system, where electron-electron 
interactions are screened. The Hamiltonian can be written 
in a single particle basis as 
\begin{equation}
H = \sum_i \epsilon_i | i \rangle \langle i | + 
\sum_{\langle i,j \rangle} t_{ij} |i\rangle \langle j |,
\label{eq:ham-def}
\end{equation}
where  $\epsilon_i$, and  $t_{ij}$ are random site energies
and  the nearest neighbour hopping energies.  
In this work, the sites energies are considered 
Gaussian distributed for each lattice system and the hopping energies are kept constant: $t_{ij}=-1.0$. 
The  Gaussian type site disorder has  been widely 
studied  for the square lattice system and has a close analogy with  
real thin film materials.

{\it The ensemble}: 
The presence of the disorder in the system makes it inevitable to consider 
an ensemble of the Hamiltonians. 
For the Hamiltonian in Eq.(\ref{eq:ham-def}) with on-site random energies, the probability 
density $\rho(H)$ of the ensemble including all possibilities can be written in general as
\begin{eqnarray}
 \rho (H) &=&
 C \, {\rm exp}\left[{-\sum_{k} {1 \over 2 h_{kk}} (H_{kk}-b_{kk})^2 } \right] 
 \nonumber \\
& & \times \prod_{k,l; k\not=l} \delta(H_{kl} -f_{kl}),
\label{bde1}
\end{eqnarray}
where $C$ is the normalization constant, $f_{kl}=t$ if $\{k,l \}$ correspond 
to nearest neighbour sites and otherwise $f_{kl}=0$. Writing the  delta function as a limiting Gaussian 
($\delta(x) = \lim_{ v \rightarrow 0}   {1 \over \sqrt{2\pi v^2}}\;  
{\rm e}^{-x^2/2 v^2} $), reduces 
Eq.(\ref{bde1}) to the following form
\begin{equation}
\rho (H,h,b) = C \exp \left[\sum_{k \le l} \frac{1}{h_{kl}} 
(H_{kl} - b_{kl})^2 \right],
\label{eq:rho}
\end{equation}
where $h$ is the set of all variances $h_{kl} = \langle H_{kl}^2 \rangle - 
\langle H_{kl} \rangle^2$, and $b$ is the set of all mean values 
$b_{kl} = \langle H_{kl} \rangle$. 
Here, for constant hopping between nearest neighbor sites and an onsite Gaussian disorder $W$ in Eq.(\ref{eq:ham-def}), one has 
\begin{eqnarray}
h_{kk} &=& W^2/12, \qquad b_{kk} = 0, \\
h_{kl} &=& 0, \qquad b_{kl} = -f(kl) t,
\label{and-pm}
\end{eqnarray}
where $f(kl) = 1$ for $\{k,l\}$ pairs representing nearest 
neighbors and zero otherwise.
The statistical behavior of the Hamiltonian $H$, Eq.(\ref{eq:ham-def}), can now be analyzed using  the ensemble (\ref{eq:rho}). An important aspect of Eq.(\ref{eq:rho}) is that it  can represent the ensemble density for a wider class of lattice systems under different  
conditions e.g., random, anisotropic, variable range hopping, dimensionality and boundary conditions; the only constraint on these system is to  preserve the time-reversal symmetry 
which allows $H$ to be real-symmetric. Its form is therefore appropriate   
for the verification of single parameter scaling conjecture.

{\it Single parameter formulation}:
Before proceeding to numerical analysis,  it is helpful to briefly review 
the complexity parametric formulation of the spectral statistics.
The variation of system conditions in general  result in a variation of the distribution parameters $h$ and $b$ and therefore an evolution of the $H$-ensemble. As discussed in  Ref.~\onlinecite{shukla05,shukla2005pre}, under a change of parameters $h_{kl} \to h_{kl} + \delta h_{kl}$ and $b_{kl} \to b_{kl} + \delta b_{kl}$, $\rho(H)$ undergoes a 
diffusion dynamics along with a finite drift, and, using Gaussian 
nature of $\rho(H)$, it can exactly be shown that
\begin{equation}
T \rho = L \rho,
\label{eq:diff}
\end{equation}
where $T$ is a combination of parametric derivatives, and $L$ 
is a diffusion operator in matrix space and are given by
\begin{equation}
T = \sum_{k \le l} \left[(g_{kl} - 2 h_{kl} ) 
\frac{\partial}{\partial h_{kl}} - \, b_{kl} \, 
\frac{\partial}{\partial b_{kl}} \right]
\end{equation}
and 
\begin{equation}
L = \sum_{kl} \frac{\partial}{\partial H_{kl}} \left[ \frac{g_{kl}}{2}
\frac{\partial}{\partial H_{kl}} +  \, H_{kl} \right]
\end{equation}
with $g_{kl} = 1 + \delta_{kl}$. A suitable 
transformation of parametric space maps $T$ to a single 
parametric derivative, $T \rho = \frac{\partial \rho}{\partial Y}$, 
which in turn reduces Eq.~\eqref{eq:diff} to a single 
parametric diffusion equation~\cite{shukla05, shukla2005pre}
\begin{equation}
\frac{\partial \rho}{\partial Y} = L \rho.
\label{diff-rho}
\end{equation}
Here
\begin{equation}
Y = \frac{-1}{N^2} \sum_{k \le l}  \left[
  \;  {\rm ln} \left| 1-   (2-\delta_{kl}) h_{kl}  \right| \;  +
\;  {\rm ln} \left| b_{kl}+\delta_{b0} \right|^2 \right] + C_y 
\label{eq:sp1}
\end{equation}
with $C_y$ as an arbitrary constant and $\delta_{b0} = 1$ 
if $b_{kl} = 0$ else $\delta_{b0} = 0$. 
For the ensemble described by the set of parameters in Eq.(\ref{and-pm}), 
this leads to
\begin{eqnarray}
Y = \frac{-1}{ N} \; {\rm ln} \left[\left|1-\frac{W^2}{12}\right|  .  \left|t + \delta_{t0}  \right|^{z/2} \right]+C_y,
\label{ya}
\end{eqnarray} 
where $z$ is the coordination number of the lattices.

The joint probability distribution of the eigenvalues $P(\{E_n\}) \equiv P(E_1, \ldots, E_N)$ for a metal (fully extended eigenfunctions limit) is given by the Wigner-Dyson distribution, $P(\{E_n\}) = \prod_{i<j} |E_i - E_j| 
\exp(-\frac{1}{2} \sum_{k} E_k^2)$. The eigenvalues are uncorrelated  in insulator  limit, thus implying   $P(\{E_n\}) = \prod_{i<j} P_n(E_n)$ \cite{shklovskii93}. The distribution for the intermediate states of localization can be  derived by  integrating $\rho$ over the associated eigenvector space. As shown in Ref.~\onlinecite{shukla05}, an integration of Eq.(\ref{diff-rho}) leads to a  single parametric diffusion of the eigenvalue distribution of the ensemble (\ref{eq:rho})  
\begin{equation}
\frac{\partial P}{\partial Y} = \sum_n \frac{\partial}{\partial E_n}
\left[\frac{\partial}{\partial E_n} + \sum_{m \neq n} 
\frac{1}{E_m - E_n} +  E_n \right] P.
\label{diff-eig}
\end{equation}

The above equation can be used to obtain the ensemble averaged level density as well as its local fluctuations  \cite{shukla05}.  As discussed in Ref.~\onlinecite{pandey95}, while the diffusion of the average level density is governed by $Y$, the diffusion of its fluctuations occurs at a scale determined 
by $(Y - Y_0) \sim \Delta_l^2$ where $Y_0$ is the value of $Y$ at 
the beginning of evolution  and $\Delta_l$ is 
the local mean level spacing: 
\begin{eqnarray}
\Delta_{local}(E) \approx L^d \; \xi^{-d} \; \Delta(E)
\label{del}
\end{eqnarray}
with $\xi$ as the localization length and $d$ as the dimension (here $d=2$). 
The statistics other than mean level density is therefore governed by 
a rescaled parameter $\Lambda(E)$:
\begin{equation}
\Lambda (E) = \frac{|Y-Y_0|}{\Delta_l^2}.
\label{eq:sp2}
\end{equation}
The solution of Eq.(\ref{diff-eig})  at $\Lambda \to \infty$ corresponds to the Wigner-Dyson statistics. The $\Lambda \to 0$ limit correspond to the distribution at the initial state of the evolution; for an insulator  initial state,  the spectral statistics reduces to the statistics of uncorrelated energy levels. 

The transition parameter $\Lambda$ is in general a function of various parameters e.g., disorder, system size, dimensionality, energy range of interest, lattice topology. Although both $Y$ and $\Delta_l$ contribute to the system dependence of $\Lambda$, the crucial influence comes from $\Delta_l$ due to its dependence on the localization length $\xi$. For finite system sizes $N$, a variation of system conditions e.g., disorder leads to a smooth crossover of statistics between the stationary limits $\Lambda \to 0$ and $\Lambda \to  \infty$. In infinite size limit, the statistics abruptly changes from  one stationary limit to the other.  If however, the limit $\Lambda^*=\lim_{N \rightarrow \infty} \Lambda(N)$ exists, the corresponding statistics would belong to a universality class different from the two stationary limits. The existence of $\Lambda^*$ therefore is a criteria for the existence of critical spectral statistics \cite{shukla05}. 

\section{\label{sec:fluc}Fluctuation measures: the definitions} 

Our main objectives in this paper is to study the influence of the system conditions on the statistical behavior and 
identify the critical regime. For this purpose, we consider four different fluctuation measures namely,
density of states (DOS), reduced partition number ($P/L$), the peak position 
of the NNSD and  the dc electrical conductivity which can briefly be described as follows.

{\it Spectral measures}: A well-known measure to analyse the short range correlations among energy levels is the nearest neighbour spacing distribution (NNSD)  which describes the probability $P(s)$ of two nearest neighbour energy levels to be found at a distance $s$  measured in the units of the mean level spacing around the desired energy regime. For the weak disorder regime with extended  eigenfunctions,  $P(s)$ is given by the Wigner surmise 
\begin{equation}
P_W(s) = \frac{\pi}{2} s \exp 
\left(-\frac{\pi}{4} \; s^2\right).
\label{pwig}
\end{equation}
For the opposite limit of  strong disorder with fully localized wavefunctions, $P(s)$ follows Poisson distribution
\begin{equation}
P_P(s) =  \exp \left(-s \right).
\label{ppoi}
\end{equation}
To compare the level spacing distribution for  the entire transition within an arbitrary energy regime for different lattices, we use a traditional measure, namely, the cumulative nearest neighbor spacing distribution $\eta_i$ which 
depends on the tail behaviour of the nearest  neighbour spacing distribution and is defined as 
\begin{equation}
\eta_i = \frac{\int_{s_i}^{\infty} P(s,\Lambda)ds - 
\int_{s_i}^{\infty} P_W(s)ds}{\int_{s_i}^{\infty} P_P(s)ds 
- \int_{s_i}^{\infty} P_W(s)ds},
\label{eq:eta}
\end{equation}
where $s_i$, $i=1,2$ refer to the two crossing points of $P_W(s)$ and $P_P(s)$: $s_1 = 0.473$ and $s_2 = 2.002$ \cite{shklovskii93,berkovits98,cuevas99}.  As the system makes a transition from delocalized to localized state, 
$\eta_i$ changes from 0 to 1. While NNSD gives the short range correlations of the energy levels, there is 
another measure which gives the long range correlations, namely, the number 
variance. It is defined as the variance of the number $n$ of unfolded energy 
levels in an energy interval $r$ centered at the energy regime of interest
i.e., the number variance 
$\Sigma^2(E,r) = \langle (n(E,r) - \langle n(E,r) \rangle)^2 \rangle$.

{\it Participation number}: The dependence of $\Lambda$ on the localization length  $\xi$ through 
$\Delta_l$ results in sensitivity of the energy level statistics to 
eigenfunction behaviour. This motivates to analyse a standard measure 
for the eigenfunction localization, namely, the reduced participation 
number, referred here as $P/L$, which characterize the spread of 
eigenfunctions in the site basis. The partition number $P$ for a 
wavefunction $\psi_n$ is defined as $P^{-1} = \sum_i^N |\psi_{in}|^4$
and is proportional to the localization length. 

{\it DC conductivity}: The localization of electronic wave-function can be characterized  
by the DC electrical conductivity $\sigma$ which can be measured also 
experimentally for real systems. Here we numerically calculate  
$\sigma$ using Kubo-Greenwood formula (see Ref.~\onlinecite{niizeki79}
for details)
\begin{equation}
\sigma(E_F) = \frac{2 \pi \hbar}{\Omega}{\rm{Tr}} \left[ \bm{J}
\delta(E_F \bm{I} - \bm{H}) \bm{J} \delta(E_F \bm{I} - \bm{H})\right]
\end{equation}
where $E_F$ is the Fermi energy, $\bm{H}$ is the Hamiltonian, 
$\Omega$ is the volume of the system  and 
$\bm{J}$ is the one electron current operator 
\begin{equation}
\vec{\bm{J}} = \frac{i e V}{\hbar} \sum_{\langle i, j \rangle} 
(\bm{R}_i - \bm{R}_j)(|i\rangle \langle j | - | j \rangle \langle i | )
\end{equation}
with $\bm{R}_i$ as the position vector of site $i$.

\section{\label{sec:res} Numerical analysis and Results }

To study the influence of system conditions on the statistical behavior, we apply 
exact diagonalization technique to numerically obtain
the eigenvalues and the eigenfunctions of the Hamiltonian (\ref{eq:ham-def})  for four different lattice types.  For each lattice type, an ensemble of approximately $1000$ realizations is considered to attain statistical accuracy. To explore the size-dependence of the statistics and its sensitivity to the energy range, many system sizes,  varying from
$L = 40$ to $L=100$, are analysed for each lattice type and for two energy ranges. The latter correspond to two filling of the Hamiltonian: firstly,  the half-filling which is a more natural choice of the systems like graphene, gold etc., and  secondly, 
a filling where the  density of states is significantly high, making a 
rigorous statistical analysis of energy levels more feasible. For the local fluctuations analysis, it is necessary to first rescale the energy eigenvalues by the local mean level spacing \cite{haake2010} (also known as {\it unfolding} of the levels).  The average density of states for this purpose is calculated using the binning method. To analyse the local spectral fluctuations  in the desired energy-range,  approximately $\sim 3\%$ of the rescaled energy  levels  are taken around the chosen Fermi energy.
The rescaled energy levels and normalized eigenfunctions are  then used to calculate the 
density of states (DOS), reduced partition number ($P/L$), the peak position 
of the NNSD and  the dc electrical conductivity. The results are arranged in order of increase in  
the coordination numbers in the  subsections A, B, C and D 
containing honeycomb, kagom\'e, square and triangular lattices 
respectively. Here we describe three different  methods of calculation in 
detail for the honeycomb lattice only;  the calculations 
for other lattice systems are carried out following similar methods.

\subsection{\label{sub:hc}Honeycomb lattice}  
The 2D honeycomb structure  is one of the most interesting lattice 
types   which occurs in many systems of industrial importance e.g., graphene. 
In this case, the non-interacting electronic  models like uniform Anderson model  
or tight binding model have two unique Dirac cones in a brillouin zone (BZ). 
The DOS $\rho(\epsilon)$ for a  2D honeycomb lattice in the clean limit (i.e. absence of disorder) is given by 
\begin{equation}
\rho(\epsilon) = -\frac{1}{\pi} \text{Im} \left[\lim_{\Delta \epsilon \to 0^+} 
   \int_{1^{\text{st}}\text{BZ}} d{\bf{k}}
   \frac{V}{V^2 - |\mu|^2 t^2} \right],
\end{equation} 
where $V = \epsilon + i \Delta \epsilon$ and $\mu = \exp(i k_x a) + 
 \exp[i(-k_x a/2 + k_y \sqrt{3} a/2)] + 
 \exp[i(-k_x a/2 - k_y \sqrt{3} a/2)]$ with $k_x$ and $k_y$ as the
components of the wave vector \cite{mckinnon93}; 
$\rho(\epsilon)$ in this case vanishes at $\epsilon=\pm 3t$. 
In half-filled limit, the system shows a gapless state and a linear 
dispersion relation which can be mapped to that of a massless Dirac state. 
In real materials, the vacancies can lead to coupling of cones but coupling 
strength remains very small,  the two cones being separated by a large momentum 
vector \cite{suzuura2002}. The microscopic calculation for the two valley Hamiltonian indicates a crossover 
from anti-localization to  weak localization \cite{suzuura2002}. A recent experiment 
also confirms the anti-localization  and weak localization of states at low and high carrier 
concentration respectively (for defects like charge impurity etc.). 
    
\begin{figure}[tbp]
\includegraphics[width=0.9\columnwidth]{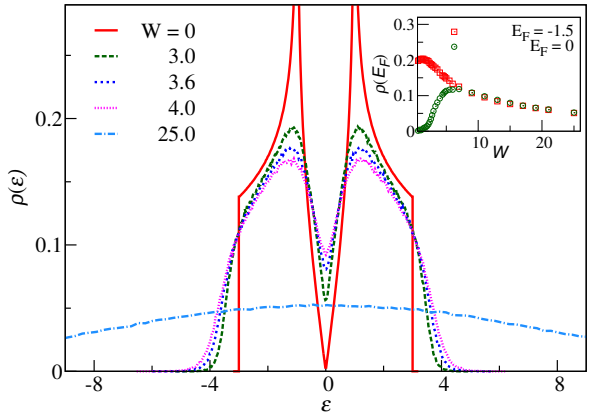}
\caption{\label{fig:doshex}(Color online) The DOS of  the honeycomb lattice. 
The solid line shows the DOS in the absence of disorder.
Three disorders strength are chosen around the critical disorder 
(see table \ref{tab:hex}). The DOS at very large disorder is also shown.
The inset shows the variation of DOS with disorder at two energy regimes.}
\end{figure}

To observe the effect of disorder, 
the DOS $\rho(\epsilon)$ is calculated numerically for different disorders; the results 
are shown in Fig.~\ref{fig:doshex}. In the clean system ($W=0$), $\rho(\epsilon)$ 
is vanishingly small near energy $\epsilon=0$ (the Fermi-energy at half filling) 
and has van-Hove singularities at $\epsilon=\pm t$ . The effect of varying disorder 
on $\rho(\epsilon)$ at $\epsilon=-1.5 t$ (i.e., at the bulk of DOS)  and at 
$\epsilon = 0$ (i.e., at the half filling) is shown in the inset of 
Fig.~\ref{fig:doshex}. At the half filled energy regime, $\rho(\epsilon)$
increases till $W=7.0 t$ and then decrease exponentially as shown 
in the inset of the Fig.~\ref{fig:doshex}. In large $W$ limit, DOS
becomes flat as shown in the Fig.~\ref{fig:doshex}. As clear from 
the Fig.~\ref{fig:doshex}, the disorder  has a significant 
impact on the DOS of this lattice type in the energy regime near $\epsilon=0$ 
(with DOS showing a prominent dip for small disorder) and $\epsilon=\pm 1.5 t$. 
For the fluctuations analysis  we therefore, choose Fermi energy from these
two regimes, namely,  $\epsilon=0$ (half filling) and $\epsilon=-1.5 t$ (bulk). 
\begin{figure}[tbp]
\includegraphics[width=0.9\columnwidth]{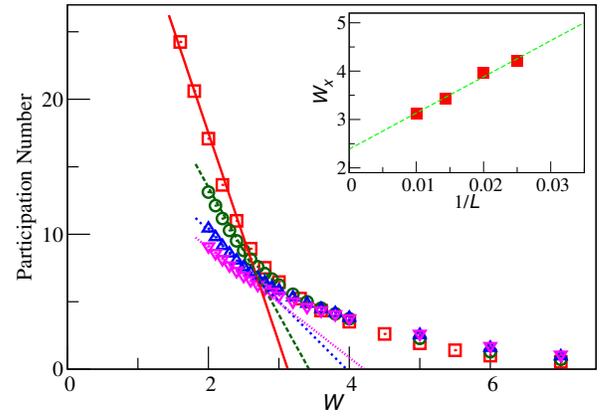}
\caption{\label{fig:part-hex}(Color online) The partition number per unit length 
with disorder for different system sizes are shown at (a) $E_F = -1.5 t$
and (b) $E_F = 0$. The linear fits of the curves cut the disorder axis at
$W_x$. The insets of both (a) and (b) show the critical disorder
$W_x$ with $1/L$ which are  fitted with the lines: 
$W_x = 3.13 + 68.6/L$ for inset (a) and $W_x=2.64 + 72.6/L$ for inset (b).}
\end{figure}
\begin{figure}[tbp]
\includegraphics[width=0.9\columnwidth]{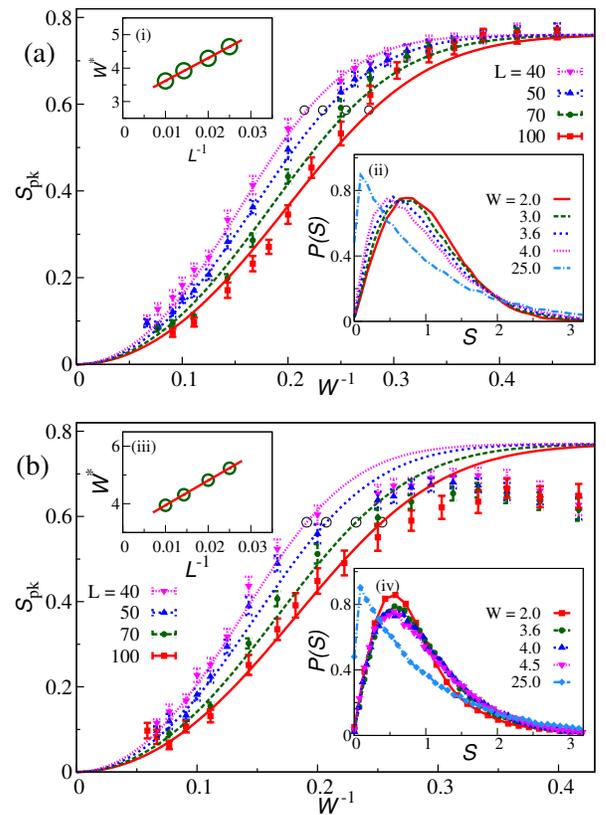}
\caption{\label{fig:pkpos-hex}(Color online) The peak-position of the NNSD with 
the inverse of the disorder (a) at $E_F = -1.5 t$ and (b) at 
$E_F = 0$. The lines in the main part of (a) and (b) are the 
fitted curves with the function $0.77 \tanh((W^*/W)^2)$. The 
critical disorder for various system sizes are depicted in 
insets (i) and (iii). The insets (ii) and (iv)  show the NNSD 
at different disorders for the system size $L = 100$}
\end{figure}
\begin{figure}[tbp]
\includegraphics[width=0.9\columnwidth]{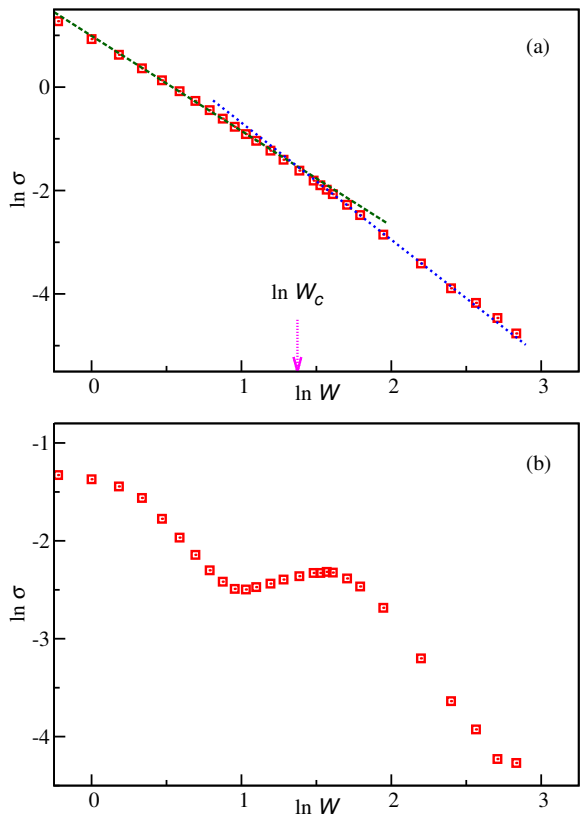}
\caption{\label{fig:cond-hex}(Color online) The dc conductivity with disorder
at (a) $\epsilon = -1.5 t$ and (b) $\epsilon = 0$. The upper part
in (b) is fitted with: $\ln \sigma = 0.99 - 1.84 \ln W$ and the lower
part is fitted with $\ln \sigma = 1.577 - 2.265 \ln W$ which gives
$W_c = 3.9 t$.}
\end{figure}

Our next step is to seek  the critical disorder for the extended to localized 
state transition in the honeycomb lattice by three different routes. While 
the transition in principle takes place in the thermodynamics limit $L \rightarrow \infty$ at a critical disorder (or critical energy),  the finite systems undergo a smooth crossover within a critical regime (around critical disorder or energy). It is therefore 
imperative to analyze the critical disorder $W_{crit}(L)$ for 
many finite system sizes,  with exact critical disorder 
given by $\lim_{ L \rightarrow \infty} W_{crit}(L) = W_{crit}$. For this purpose, 
we first  analyse the disorder dependence of the reduced 
participation number $P/L$. Fig.~\ref{fig:part-hex} elucidates  
the behavior of the reduced partition number at two Fermi energies
at (a) the bulk  and (b) at the half filling respectively. For each 
case $P/L$ varies linearly with the disorder $W$ in the small disorder 
regime while it varies exponentially for strong disorders.
The linear portion of the curves are fitted with lines ($P/L = a - b W$) 
which cut the disorder axis at $W_x = a/b$. For $W > W_x$, 
the localization length of the systems is less than the 
system size. At $W=W_x$, the localization length become equal 
to the system size $L$, thus suggesting $W_x$ as the  finite size critical disorder 
(more clearly $W^{pn}_{crit}(L)=W_x$ with superscript referring to the method applied) . 
(Note,  for later 
reference, here we use different notations for the critical disorder obtained 
by different methods).
The critical disorder measured from this method for different system 
sizes of the honeycomb lattice are  given in the table~\ref{tab:hex}.
As shown in the  insets of Fig.~\ref{fig:part-hex}(a) and (b),  
$W_x$ for different system sizes vary linearly with the inverse of 
the system size. The numerics reveals the sensitivity of $W_x$  to 
Fermi energy too: for $L=100$, the value of $W_x$ is 3.4 at the half 
filling and 3.8 at the bulk of the DOS.

Continuing with our quest for critical disorder, our next step is to analyze the 
Nearest neighbor level spacing distribution (NNSD) defined in section III.  
The decrease of disorder in the finite systems causes the NNSD to crossover from 
the Poisson distribution (eq.(\ref{ppoi}))  to the Wigner surmise (eq.(\ref{pwig})). To analyse 
the NNSD for different disorder strengths, the energy eigenvalues 
are apriori unfolded using the local mean level spacing.  The numerically obtained 
NNSD for a given disorder and system size is compared with the Brody distribution \cite{brody81}:
\begin{equation}
P_B(S) = \alpha (1+\omega) S^{\omega} \exp(-b S^{1 + \omega})
\label{brody}
\end{equation}
which gives the Brody parameters $\omega$ and $b$ as a function of $W$ and $L$. The fitted Brody distribution is then 
used to calculate the peak position dependence on the  disorder strengths $W$ for a 
fixed size $L$.  
The dependence is plotted with respect to inverse of the disorder in 
Fig.~\ref{fig:pkpos-hex} for many system sizes and at two Fermi energies 
(the bulk and the half filling).
The curve depicting peak positions with respect to disorder for each $L$ 
is now fitted with the function 
\begin{eqnarray}
F(W)=0.77 \tanh((W^*/W)^n)
\label{fw}
\end{eqnarray} 
where $n=2$ and $W^*$ is the critical disorder strength (in this method) at which localization occurs for a finite system size $L$ (more clearly $W^{nnsd}_{crit}(L)=W^*)$. 
This type of functional 
dependence of the peak position of NNSD was suggested by A J Millis {\it et al.}  
in context of the energy level statistics of one dimensional spin-1/2 
chain to find the critical value  $J_c$ indicating  integrable 
and non-integrable  boundary \cite{millis2004}.  
As shown in the main Fig.~\ref{fig:pkpos-hex} (b), 
there is large deviation in the peak position-disorder curve from the 
function $F(w)$ in the small disorder regime for the half filled 
case.  $F(W)$ with $n=2$ however fits well in small $W$ limit 
for the case where the Fermi energy is chosen away from 
the half filling. As  shown  in the insets (i) and (iii) of 
Fig. \ref{fig:pkpos-hex} (a) and (b) respectively,
the crossover disorder $W^*$ decreases with system size 
$L$.  Again  $W^*$ shows an energy dependence for a fixed $L$: 
for  $L=100$, $W^*=3.9 $ and $3.6$ for $\epsilon=0$  and 
$\epsilon= -1.5 t$ respectively (see table \ref{tab:hex}). 
The insets (ii) and (iv) 
of Fig. \ref{fig:pkpos-hex} (a)  and (b) also display the NNSD behavior 
for $\epsilon=0$ and $\epsilon=-1.5 t$ respectively.  
We notice that for $\epsilon=0$ or half filled case the
NNSD vary significantly  in small impurity limit  but the 
variation in the bulk limit $\epsilon =-1.5 t$  is weak below 
$W=W^*$. The unusual behaviour can be explained from the rapid 
change in the DOS with disorder at the half filling. The half filling
case suggests that the system is still in the ballistic limit for 
small disorder.

The electronic conductivity is an important characteristic of the localization to delocalization transition, with  high conductivity an indicator of the large localization length. The dc electrical conductivity for various system sizes for the two energy ranges  
are calculated using the Kubo-Greenwood formula. The minimum
conductivity $\sigma_m $ of a clean sample at $T=0$ and half-filled limit
 is analytically predicted to be $4e^2/\pi h$; this is also confirmed by our analysis. 
The effect of changing disorder on the behavior of   conductivity ($\sigma$)  
in honeycomb lattice is displayed in the figure \ref{fig:cond-hex}
(a) and (b) at the bulk of DOS  and at the half filling respectively. 
In the bulk limit, the conductivity decreases 
with increasing disorder following a power law dependence with exponents $-1.84$ 
in the weak disorder regime and $-2.265$ in the strong disorder regime. 
The intersection of these two lines gives the crossover
disorder $W^{cond.}_{crit}(L)= W_c = 3.9 t$ which is almost in agreement with the values 
calculated from the other two methods. The disorder-dependence of 
conductivity at the half filling however shows some atypical behavior:  
it remains almost constant at very small disorders up to $W=0.4 t$, decreases 
thereafter up to  $W<2.8t$, then increases till $W=5.0$ and  decrease 
afterwards for further increase in $W$. A possible explanation of this 
flip-flop behavior may lie in poor  statistics due to small number of states 
available at this filling or due to  dominant finite size effect. We   intend to 
probe this behavior with more numerical rigor separately.  
\begin{table}[t]
\begin{center}
\setlength{\tabcolsep}{1.0em}
{\renewcommand{\arraystretch}{1.2}
\begin{tabular}{ c  c  c  c  c } \hline
\hline
$\epsilon$ & $L$ & $W_x$ & $W*$ & $W_c$ \\ \hline
\multirow{4}{*}{$-1.5 t$} & 40  & 4.9 & 5.2  \\
         & 50  & 4.4 & 4.8 \\
         & 70  & 4.2 & 4.3  \\
         & 100 & 3.8 & 3.9 & 3.9 \\
\hline
\hline
\end{tabular}
}
\end{center}

\caption{\label{tab:hex}Critical disorders calculated from all three 
methods for the honeycomb lattice at two different Fermi energies for
four system sizes}
\end{table}

Next we compare the crossover values of $W_{crit}(L)$ calculated from 
three different methods (i.e $W_x, W^*, W_c$) for four different $L$. Due to statistical 
reliability, the comparison is shown in table \ref{tab:hex} (only for the 
bulk energy limit). As given in the table \ref{tab:hex}, $W_x$ and $W^*$ decrease with an  
increase in the system size $L$, varying as $L^{-1}$ 
(As the size dependence on the conductivity is negligibly small for this case, 
we present the data for $L = 100$ case only).
This clearly indicates  
conductivity as a better measure to find the critical disorder for the transition (due to faster convergence with $L$). The analysis suggests the critical disorder $W_{crit} \approx 3.8$.

\subsection{ Kagom\'e lattice} 
\begin{figure}[t]
\includegraphics[width=0.9\columnwidth]{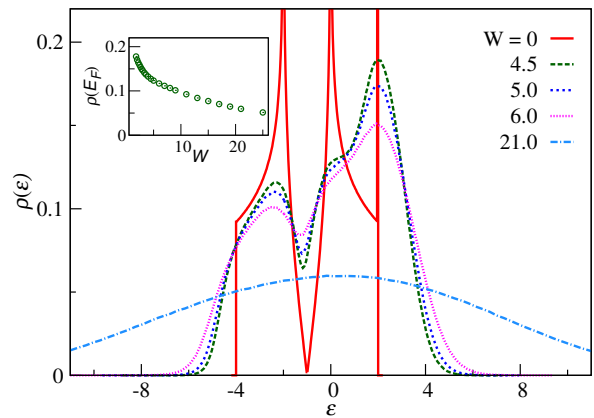}
\caption{\label{fig:doskag}(Color online) The DOS of the kagom\'e lattice
for different disorders. The solid line is the DOS for the 
clean system. The inset shows the DOS at half filling with
the disorder.}
\end{figure}

The DOS $\rho(\epsilon)$ of the Anderson model with uniform site energy for a 
kagom\'e lattice is exactly solvable in the clean limit, and is 
composed of the two bands of the honeycomb lattice shifted 
in energy by $t$ with amplitude reduced by a factor of 2/3
and a delta peak at $\epsilon = 2 t$ with weight 1/3.  The DOS 
diverges at three values of energy (at $\epsilon = 0$ and $\pm 2 t$) and is 
zero for $\epsilon=-t$ for the clean system \cite{hanisch95}. The 
band width in the clean limit of this lattice is spanned 
from $\epsilon = -4 t$ to $\epsilon = 2t$. Effect of the onsite 
disorder on the DOS of the kagom\'e lattice Hamiltonian  of system 
size $L^2$ with  $ L \sim 100 $ is shown in the 
Fig.~\ref{fig:doskag} for three disorder strengths alongwith the clean limit. 
As clear from the figure, the singularities in the DOS vanish with the increase in 
disorder, with DOS approaching a Gaussian distribution in  energy
at very large disorder. The inset in Fig~\ref{fig:doskag}  displays the 
behavior of the DOS at the Fermi energy for varying disorders which turns out to be 
an exponential decay in large disorder limit. 
\begin{figure}[b]
\includegraphics[width=0.9\columnwidth]{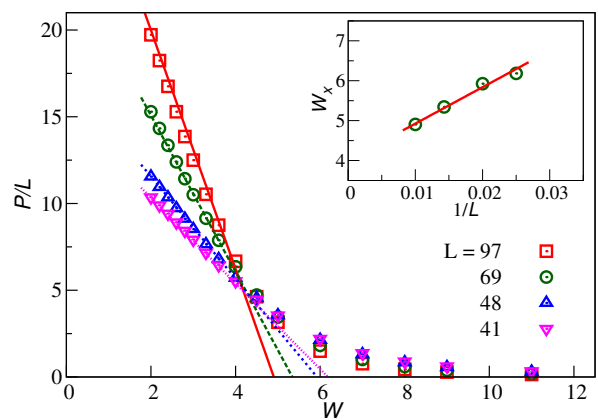}
\caption{\label{fig:partnum-kag}(Color online) The reduced partition number
for different system sizes of the kagom\'e lattice. The linear 
portion of the curves are fitted with lines which cut the 
disorder axis at $W_x$. The inset shows the critical disorder 
$W_x$ with the inverse of the system size and which follows 
the fitted line: $W_x = 4.0 + 91.3/L$.}
\end{figure}
\begin{figure}[t]
\includegraphics[width=0.9\columnwidth]{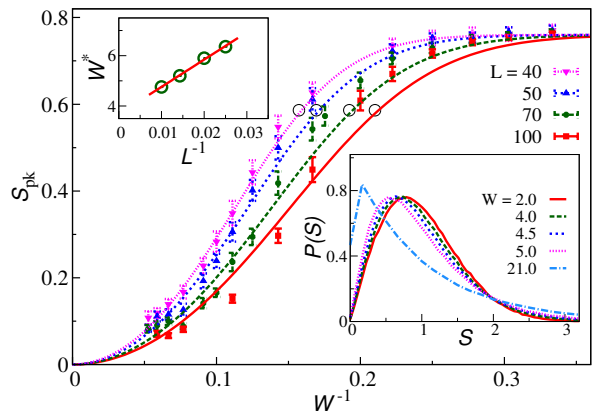}
\caption{\label{fig:pkpos-kag}(Color online) The peak-position of the NNSD 
vs the inverse of the disorder. The lines in the main
figure are  fitted curves with $0.77 \tanh((W^*/W)^2)$. 
The top left inset shows the critical disorder with 
the system size which follows: $W^* = 3.67 + 108.5/L$. 
The bottom right inset shows the NNSD of the kagom\'e 
lattice of system size $L \sim 100$ for different disorders}
\end{figure}
\begin{figure}[b]
\includegraphics[width=0.9\columnwidth]{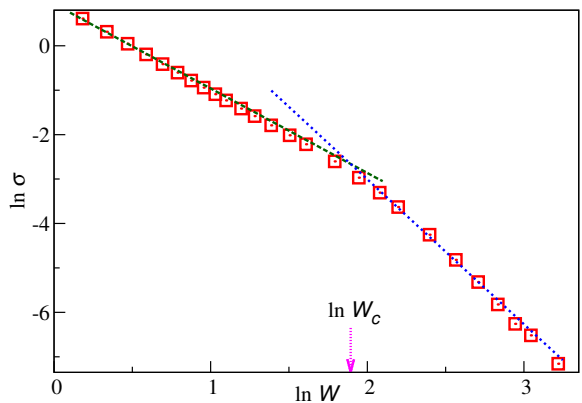}
\caption{\label{fig:cond-kag}(Color online) DC conductivity of the kagom\'e 
lattice of linear size $\approx$ 100 with disorder. The upper
part of the plot follows: $\ln \sigma = 0.938 - 1.9 \ln W$ and 
the lower part follows: $\ln \sigma = 3.514 - 3.2 \ln W$ and 
the critical disorder $W_c = 5.9$.}
\end{figure}

To determine the critical disorder $W_{crit}$ in this case, we again apply the three 
methods mentioned in detail for the honeycomb 
lattice. The results for reduced participation number $P/L$  
are displayed in  Fig.~\ref{fig:partnum-kag} for different system sizes in the kagom\'e 
lattice. As shown in the  inset of 
Fig.~\ref{fig:partnum-kag}, the finite size critical disorder $W_x$ calculated
from this method varies linearly with the inverse of the system size.
The results for the search of critical disorder through the NNSD peak position analysis 
are displayed in Fig. \ref{fig:pkpos-kag},  for four different system sizes of kogom\'e lattice. 
Here again the curve describing peak position-disorder dependence is fitted with the function $F(W)$ given by Eq.(\ref{fw})  with $n=2$. 
As  shown  in the inset 
of Fig.~\ref{fig:pkpos-kag}, $W^*$ decrease with system size $L$. The behavior of 
NNSD  shown in the inset of Fig.~\ref{fig:pkpos-kag} indicates a very small variance in NNSD 
for $W<W^*$. 

The disorder dependence  of the electronic conductivity for kagom\'e lattice is displayed in Fig. \ref{fig:cond-kag}  (shown here only for $L=100$); it reveals a decreasing conductivity with increase of disorder, along with two different power law behaviors in small and large disorder ranges (indicated by two linear regimes with different slopes in the log $\sigma$-log $W$ plot). For a system size $L \approx 100$, the crossing point of  the two lines lies at $W_c \sim 5.9t$.

Table \ref{tab:kag} displays the critical disorder $W_{crit}(L)$ calculated from three different 
methods for four different system sizes. Similar to the honeycomb lattice case; the size dependence
of $W_c$ is not presented as it is negligibly small for this system also. 
whereas $W_x$ and $W^*$ calculated from participation number and shift 
in peak height of NNSD is linear with $L^{-1}$.  
Similar to the case of honeycomb lattice, here again the calculations suggest a  
convergence of $W_c$  to its critical value  
lattice, once again confirming a weaker sensitivity of the  conductivity approach  to 
finiteness of the kagom\'e lattice.
\begin{table}[h]
\begin{center}
\setlength{\tabcolsep}{1.0em}
{\renewcommand{\arraystretch}{1.2}
\begin{tabular}{ c  c  c  c } \hline
\hline
$L$ & $W_x$ & $W*$ & $W_c$ \\ \hline
 40  & 6.2 & 6.4 & \multirow{4}{*}{5.9} \\
 50  & 5.9 & 5.9 &      \\
 70  & 5.3 & 5.2 &      \\
 100 & 4.9 & 4.8 &      \\
\hline
\hline
\end{tabular}
}
\end{center}
\caption{\label{tab:kag}Critical disorders calculated from all three
methods for the kagom\'e lattice for four system sizes}
\end{table}

\subsection{Square  lattice}  
The DOS $\rho(\epsilon)$ of an Anderson model with uniform site energy for a 
square lattice at the clean limit can be calculated from the band
dispersion relation
\begin{equation}
\epsilon ({\mathbf k})  = 2 t (\cos k_x + \cos k_y).
\end{equation}
\begin{figure}[hb]
\includegraphics[width=0.9\columnwidth]{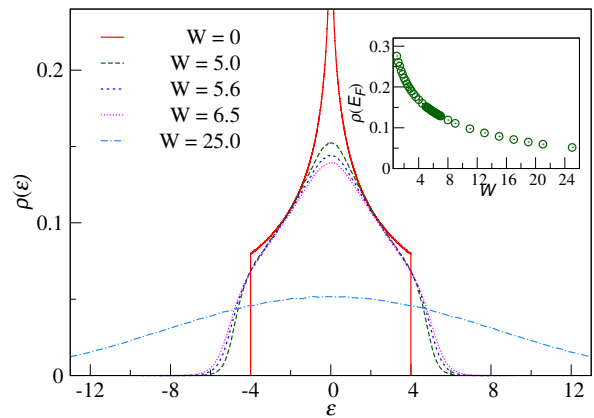}
\caption{\label{fig:dossqr}(Color online) The DOS of the square lattice at 
different disorders. The solid line is the DOS for a clean 
system. The inset shows the DOS at the half filling with
the disorder.}
\end{figure}
In the clean limit, the DOS has a van Hove singularity at $\epsilon = 0$
and vanishes for  $|\epsilon| > 4 t $.  Effect of 
on-site disorder on the DOS for four system sizes of square lattices 
is shown in the Fig.~\ref{fig:dossqr}. 
As shown in the inset of Fig~\ref{fig:dossqr} the DOS at 
half filling decays exponentially with disorder which is consistent 
with the Ref.~\onlinecite{guyu2005}. 

\begin{figure}[bp]
\includegraphics[width=0.9\columnwidth]{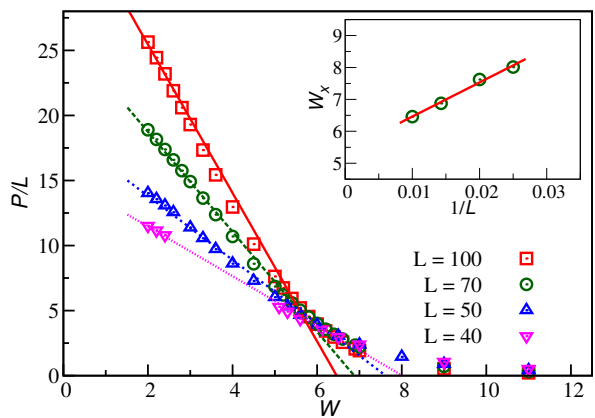}
\caption{\label{fig:partnum-sqr}(Color online) The reduced partition number of
the square lattice of different system sizes. The upper part of 
the curves are fitted with lines which cut the disorder axis at $W_x$.
The inset shows the critical disorder $W_x$ with the inverse of 
the system size and the fitted line: $W_x = 5.39 + 106.8/L$.}
\end{figure}

\begin{figure}[tp]
\includegraphics[width=0.9\columnwidth]{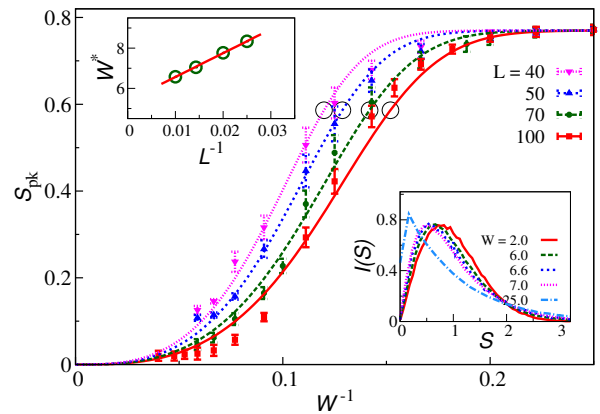}
\caption{\label{fig:pkpos-sqr}(Color online) The peak position of the NNSD 
of the square lattices of various system sizes. The curves in 
the main plot are fitted with: $0.77 \tanh ((W^*/W)^3)$. 
The top left inset $W^*$ are plotted with inverse of the system 
size which follows the line: $W^* = 5.38 + 118.4/L$. The bottom
right inset depicts the NNSD at different disorder for the square 
lattice of size $L = 100$.}
\end{figure}

\begin{figure}[h]
\includegraphics[width=0.9\columnwidth]{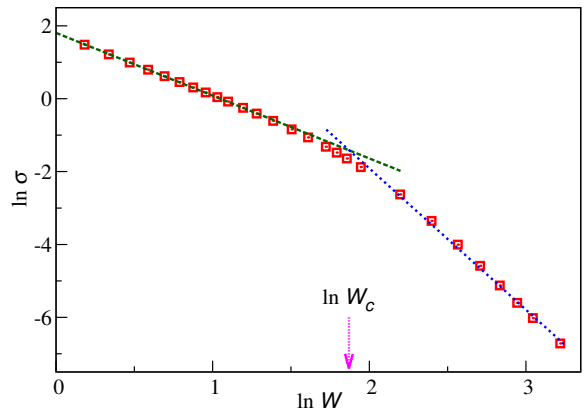}
\caption{\label{fig:cond-sqr}(Color online) DC conductivity of the disordered 
square lattice at the half filling. The upper part is fitted with:
$\ln \sigma = 1.807 - 1.723 \ln W$ and the lower part is fitted with: 
$\ln \sigma = 5.845 - 3.879 \ln W$ which gives $W_c = 6.5 t$.}
\end{figure}

Following the same procedure as mentioned in the case of honeycomb lattice,
here again we analyze the disorder dependence of the reduced partition 
number $P/L$, NNSD and 
DC conductivity  for different system sizes but only in bulk energy regime;  
the results are shown in Fig.~\ref{fig:partnum-sqr}, Fig.~\ref{fig:pkpos-sqr} and Fig.~\ref{fig:cond-sqr} respectively. As clear from the figures, the qualitative behavior 
in this case is same as in the bulk energy regimes of honeycomb and kagome lattices; 
a quantitative difference however shows up in the fitted line $W_x = 5.39 + 106.8/L$ for $P/L$, 
with power $n$ for fit $F(W)$ as $3$ (instead of $2$ as in previous two cases), 
with $W^* = 5.38 + 118.4/L$ and $W_c=6.5 t$. 
The critical disorders $W_x, W^*$ and $W_c$ 
from the three methods are given in table \ref{tab:sqr}. 
$W_x$, $W^*$ and $W_c$ for this case show an inverse linear dependence on 
the system size.

\begin{table}[!hb]
\begin{center}
\setlength{\tabcolsep}{1.0em}
{\renewcommand{\arraystretch}{1.2}
\begin{tabular}{ c  c  c  c } \hline
\hline
 $L$ & $W_x$ & $W^*$ & $W_c$ \\ \hline
40  & 8.0 & 8.3 &  \\
50  & 7.6 & 7.8 & 7.6 \\
70  & 6.9 & 7.0 &  7.0 \\
100 & 6.5 & 6.6 &  6.5 \\
\hline
\hline
\end{tabular}
}
\end{center}
\caption{\label{tab:sqr}Critical disorders calculated from all three
methods for the square lattice at half filling for four system sizes}
\end{table}

\subsection{Triangular lattice}  
\begin{figure}[tp]
\includegraphics[width=0.9\columnwidth]{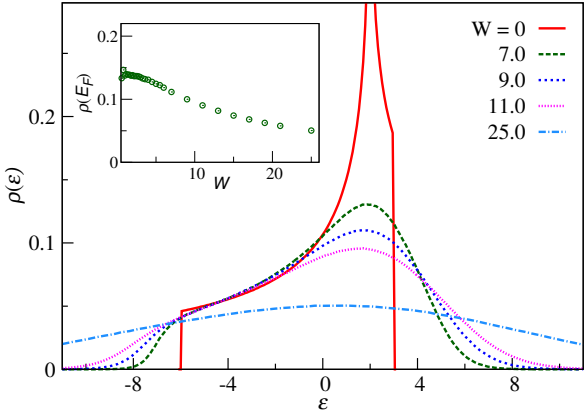}
\caption{\label{fig:dostri}(Color online) The DOS of the triangular lattice 
in presence of different disorder strengths. The solid line is
the DOS for the system without any disorder. The inset shows the
variation of DOS at the half filling with disorder.}
\end{figure}

In the clean limit, the DOS $\rho(\epsilon)$ for  a triangular  
lattice can exactly be obtained from the band dispersion 
relation 
\begin{equation}
\epsilon ({\mathbf k}) = -t \left[2 \cos(k_x) + 
   4 \cos \left(\frac{k_x}{2}\right) 
     \cos\left(\frac{\sqrt{3}}{2} k_y\right) \right],
\end{equation}
where ${\mathbf k}$ is confined to the first Brillouin Zone \cite{hanisch95}.
The DOS obtained numerically in the clean limit along with the 
presence of disorder is shown in the Fig~\ref{fig:dostri}.
The analysis indicates  a van Hove singularity in the  DOS at energy 
$E = 2t$ and  it approaches zero for $\epsilon < -6 t$ and 
$\epsilon > 3 t$. The energy dependence of the DOS
for four disorders is shown in the Fig.~\ref{fig:dostri}. 
In the inset of  Fig~\ref{fig:dostri}, the disorder dependence of the DOS at 
a fixed energy (half filled energy)  is displayed which indicates an 
exponential decrease in  DOS with increasing disorder. 

\begin{figure}[bp]
\includegraphics[width=0.9\columnwidth]{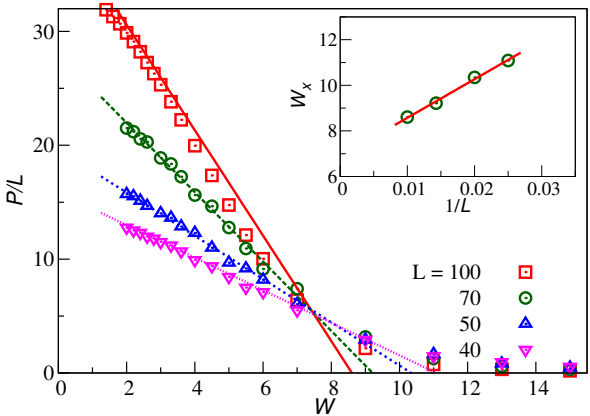}
\caption{\label{fig:partnum-tri}(Color online) The reduced partition number of 
triangular lattice of different system sizes with disorder. The 
are the linear fit for the reduced partition number curve
which cut the $W$ axis at $W_x$. The finite size dependence of 
the critical disorder $W_x$ is shown in the  inset which follows 
the line: $W_x = 6.88 + 169.3/L$.}
\end{figure}
\begin{figure}[tp]
\includegraphics[width=0.9\columnwidth]{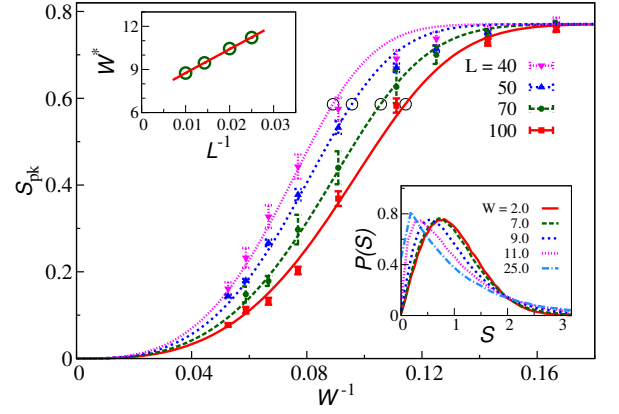}
\caption{\label{fig:pkpos-tri}(Color online) The peak position of the NNSD
with the inverse disorder for the triangular lattice of different
system sizes. The data is fitted with the function $0.77 \tanh ((W^*/W)^2)$.
The top left inset shows the variation of $W^*$ with system size 
which follows the fitted line: $W^* = 7.10 + 165.8/L$. The bottom
right inset shows the NNSD of the triangular lattice of system 
size $L = 100$ calculated at the half filling.}
\end{figure}

\begin{figure}[h]
\includegraphics[width=0.9\columnwidth]{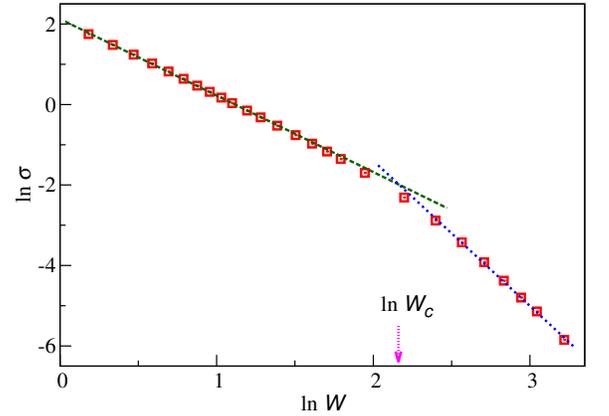}
\caption{\label{fig:cond-tri}(Color online) DC conductivity of the disordered triangular
lattice at the half filling. The upper part is fitted with: 
$\ln \sigma = 2.126 - 1.902 \ln W$ and the lower part is fitted with: 
$\ln \sigma = 5.812 - 3.606 \ln W$. The crossing point of these two 
fitted curves gives $W_c = 8.7$.}
\end{figure}

The results for the search of critical disorder in triangular lattice by the three methods 
(same as mentioned above for previous lattices) are shown in Fig.~\ref{fig:partnum-tri}, Fig.~\ref{fig:pkpos-tri} and Fig.~\ref{fig:cond-tri} respectively. The figures again indicate 
the same qualitative behavior as in the bulk energy regimes of honeycomb, kagom\'e and 
square lattices but the quantitative difference shows up in the fitted line $W_x = 6.88 + 169.3/L$ for $P/L$, with power $n$ for fit $F(W)$ as $3$ (same as in square lattice but different 
from honeycomb and kagom\'e), with $W^* = 7.10 + 165.8/L$ and $W_c=8.7 t$. 
A comparison of  critical disorders $W_x, W^*$ and $W_c$ (for Fermi energy in bulk)
from the three methods is  displayed in  table \ref{tab:tri}  which confirms, as for previous three lattice types,    
an inverse linear dependence for $W_x$, $W^*$ and $W_c$ in this case too.
\begin{table}[!hb]
\begin{center}
\setlength{\tabcolsep}{1.0em}
{\renewcommand{\arraystretch}{1.2}
\begin{tabular}{ c   c   c   c } \hline
\hline
 $L$ & $W_x$ & $W^*$ & $W_c$ \\ \hline
40  & 11.1 & 11.2  &  \\
50  &  10.3 & 10.4 & 10.4 \\
70  &  9.2 &  9.5 &  9.4\\
100 &  8.6 &  8.7 &  8.7\\
\hline
\hline
\end{tabular}
}
\end{center}
\caption{\label{tab:tri}Critical disorders calculated from all three
methods for the triangular lattice at half filling for four system sizes}
\end{table}

\section{\label{sect5}complexity parameter formulation of the transition}

As discussed in previous section, the numerical analysis reveals the qualitative 
insensitivity of the local fluctuations in  physical properties 
to system parameters (within a fixed energy range) although a quantitative 
dependence is  indicated. More clearly, for each lattice type with Fermi energy in the bulk we observe the following behavior: (i) the reduced particiapation ratio has a linear/exponential dependence on the disorder in weak/strong disorder regime, respectively, (ii)  the disorder dependence of the peak positions of NNSD can be described by the function $F(w)$ (with different $n$ values), (iii) the conductivity  has a power law dependence on the disorder with different exponents in weak and strong disorder regime. 
\begin{figure}[b]
\includegraphics[width=0.9\columnwidth]{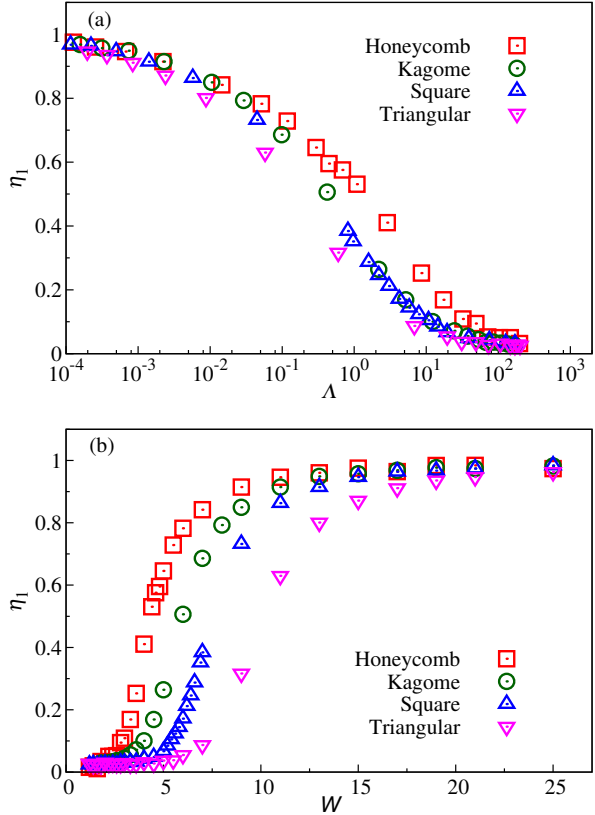}
\caption{\label{fig:eta1}(Color online) The variation of $\eta_1$ with (a) the 
rescaled complexity parameter $\Lambda$ for four different 
two dimensional lattices and (b) the disorder $W$. As obvious, 
the behaviour  for different lattices coincide in terms of 
$\Lambda$ but not in terms of disorder.}
\end{figure}
The observed behavior therefore strongly suggests the possibility of 
a common mathematical formulation of the statistical properties 
where system dependence enters through a single function of all 
system parameters.   The  theoretical steps for such a formulation 
are briefly reviewed in section II, with the function referred as 
the complexity parameter. As defined  in Eq.~\ref{ya}, the  
function $Y$ is a combination of  various system parameters, with 
explicit dependence on the disorder strength, hopping and system size.  
The information about dimensionality and boundary conditions is 
implicitly contained in the sparsity of the matrix $H$ as well 
matrix element identities and therefore in the summation in 
Eq.~\ref{eq:sp1}.  The necessary  rescaling of energy levels for 
comparison of the fluctuations however leads to $\Lambda$ 
(Eq.~\ref{eq:sp2}) as the relevant transition parameter; the 
rescaling therefore introduces the  crucial  dependence on the 
dimensionality as well as on the Fermi energy. The obvious 
relevance of the theoretically obtained single parameter 
governing the transition renders its  numerical/ experimental 
verification very desirable. For this purpose, we consider  
here three well-known spectral fluctuation measures namely, 
the cumulative NNSD $\eta_1, \eta_2$ and the number variance 
$\Sigma^2(r)$ (defined  in section III)  of the four lattices 
and analyze their evolution in terms of the complexity parameter 
$\Lambda$ instead of disorder $W$.  The verification in context 
of the eigenfunction fluctuations is yet to be carried out and 
will be reported elsewhere.  

\begin{figure}[t]
\includegraphics[width=0.9\columnwidth]{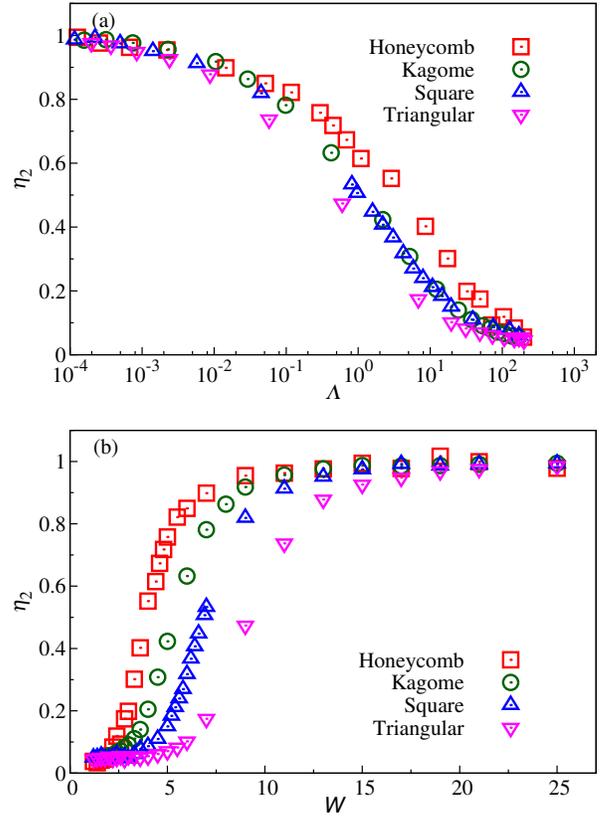}
\caption{\label{fig:eta2}(Color online) The variation of $\eta_2$ with (a) the 
rescaled complexity parameter $\Lambda$ for four different 
two dimensional lattices and (b) the disorder $W$. As obvious, 
the behaviour  for different lattices coincide in terms of 
$\Lambda$ but not in terms of disorder.}
\end{figure}
\begin{figure}[t]
\includegraphics[width=0.9\columnwidth]{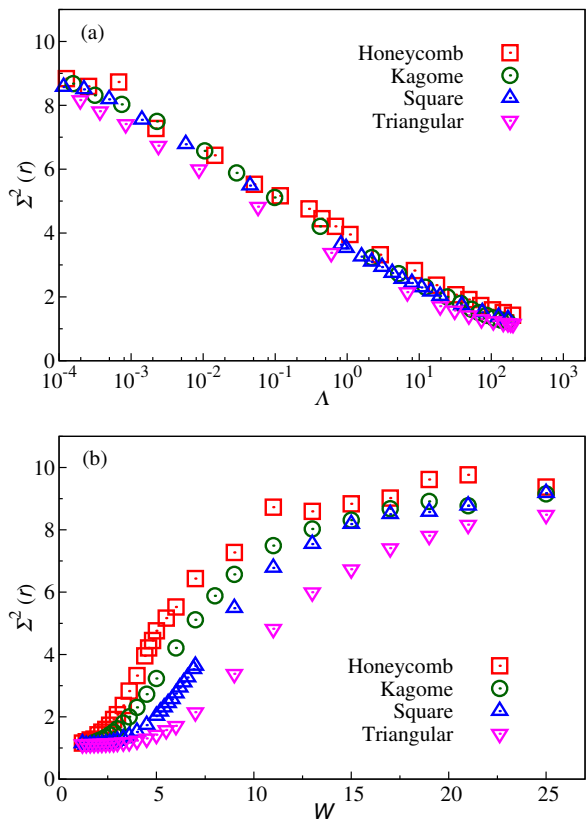}
\caption{\label{fig:nv}(Color online) The number variance $\Sigma^2(r)$ at $r = 10$ 
with (a) the rescaled complexity parameter $\Lambda$ for four different 
two dimensional lattices and (b) the disorder $W$. As obvious, 
the behaviour  for different lattices coincide in terms of 
$\Lambda$ but not in terms of disorder.}
\end{figure}

To calculate $\Lambda$ for our analysis,  we use the fact that 
the localization length $\xi$ is proportional to the average 
participation number $P$.   Fig.~\ref{fig:eta1}, \ref{fig:eta2} 
and \ref{fig:nv} show the results for $\eta_1, \eta_2$ and $\Sigma^2(r)$, 
respectively,  for four lattices for Fermi energy at $e=0$. As clear from 
the part (a) of these figures, $\Lambda$-governed evolution of each of these measure 
for all four lattices falls almost on the same curve for entire crossover from the  
localization to delocalization; the difference of  connectivity of the lattices 
does not affect their behavior. Note a disorder ($W$)-dependent evolution of 
$\eta_1$ and $\eta_2$ for honeycomb and kagome lattices is expected to differ 
from that of square and triangular ones (as suggested by the  observed difference 
in  disorder governed  evolution of the NNSD-peaks of
 the lattices, with $\eta_1$ and $\eta_2$  being cumulative NNSDs); the deviation 
of the results for four lattices is clearly visible from part (b) of  
Fig.~\ref{fig:eta1}, \ref{fig:eta2} and \ref{fig:nv}).  This clearly 
reveals $\Lambda$ as the parameter in terms of which the transition in  spectral statistics 
in Anderson lattices follows a universal route.

\section{\label{sect6}Discussion } 
From our results it is clear that the statistical behavior of two 
dimensional finite size lattices depends on the coordination 
number, the lattice connectivity  and the system size. The statistics is 
different  for the lattices with same coordination 
number with different connectivity 
which is  evident from the spectral statistics of the square lattice and 
the kagom\'e lattice (both having same coordination number). 
The spectral averaged density of states  for all the lattices, considered 
in this paper, show strong disorder sensitivity  in  weak disorder regime, and,  
at least one van-Hove singularity in clean limit. The  position of the singularity 
is sensitive to the lattice type; it occurs at energy 
$\epsilon = 0$ for the square lattice, at $\epsilon = 2.0 t$
for the triangular lattice, at $\epsilon=\pm t$ for the honeycomb
lattice and at $\epsilon=0$, $-2 t$ for the kagom\'e lattice. 
The density of states for the honeycomb lattice and the kagom\'e 
lattice are related to that of the triangular lattice \cite{hanisch95} 
and can be written in terms of the DOS of the triangular lattice
in the clean limit. The DOS at $\epsilon=\epsilon_F$ varies
differently for different lattice systems at small disorder region
whereas they all decay exponentially for strong disorder region. 
At very large disorder, the DOS for all the lattice system
lead to a Gaussian distribution.

Three methods  are used to estimate the critical disorder for 
delocalization to localization transition in two dimensional 
finite size lattices. First, the reduced partition number ($P/L$) are used to find
the critical disorder $W_x$. $P/L$ varies linearly in the weak
disorder regime whereas, it varies exponential in the strong
disorder regime.
Next, we study the peak position of the NNSD of each lattice type 
for four system sizes; which fits  with function  $0.77 \tanh((W^*/W)^{\alpha})$ 
where $\alpha = 3$ for the square and the triangular lattices
whereas, $\alpha=2$ for the honeycomb and the kagom\'e lattices. 
Last, the Kubo-Greenwood dc conductivity is to find the critical disorder ($W_c$)
as the conductivity follows two power law decay in the weak and strong
disorder regimes. The critical disorders calculated from all the methods
are in agreement with each other.

We also analyze dependence 
of the critical disorder on finite size and lattice structure considering
four system sizes for all the cases.  Our results indicate 
(i) a linear varaition of critcal disorder with $1/L$, 
(ii)  it is smallest for the honeycomb lattice and largest for the triangular lattice and 
 increases with the coordination number for a finite
lattice of size $L$. Although the coordination number is same for the kagom\'e lattice
and the square lattice the critical disorders are different for the two cases.
Therefore, the critical disorder depends not only on the coordination number 
but also on the lattice connectivity.

Finally, we compare three fluctuation measures namely, the cumulative 
NNSD measures $\eta_1$, $\eta_2$ and the number variance $\Sigma^2(r)$ for the 
four lattices and study their evolution  with the single complexity 
parameter $\Lambda$. The reults  confirm the single parameter dependence of 
the localization to delocalization transition in Anderson Hamiltonian in context of 
the spectral statistics.

\section{\label{sect7}Acknowledgments}
M. K. thanks DST for Ramanujan fellowship grant vide No. 
SERB/F/3290/2013-2014 and DST Nanomission for the  CRAY computational facility.

\end{document}